\documentclass[referee]{mn2e}
\newcommand{\be}{\begin{eqnarray}}
\newcommand{\ee}{\end{eqnarray}}

\newcommand{\Ms}{M_{\star}}

\newcommand{\Mp}{M_p}
\newcommand{\Md}{M_d}
\newcommand{\MJ}{M_{\rm J}}
\newcommand{\MN}{M_{\rm N}}
\newcommand{\nab}{\mbox{\boldmath $\nabla$}}

\usepackage{graphicx}
\usepackage{amsmath}

\graphicspath{{figures/}}

\begin{document}

\title[Orbital evolution of a planet on an inclined orbit]{Orbital
  evolution of a planet on an inclined orbit interacting with a disc}

\author[J. Teyssandier, C. Terquem and J. Papaloizou] {Jean
  Teyssandier$^{1}$\thanks{Email: teyssand@iap.fr}, Caroline
  Terquem$^{1}$\thanks{E-mail: caroline.terquem@iap.fr} and John
  C. B. Papaloizou$^{2}$\thanks{E-mail: J.C.B.Papaloizou@damtp.cam.ac.uk}\\
  $^{1}$ Institut d'Astrophysique de Paris, UPMC Univ Paris 06, CNRS,
  UMR7095, 98 bis bd Arago, F-75014, Paris, France \\
  $^{2}$ Department of Applied Mathematics and Theoretical Physics,
  University of Cambridge, Centre for Mathematical Sciences,\\
  \ \ Wilberforce Road, Cambridge, CB3 0WA, UK \\ }

\date{2012}

\pagerange{\pageref{firstpage}--\pageref{lastpage}} \pubyear{2012}

\maketitle

\label{firstpage}

%
%

\begin{abstract}
  We study the dynamics of a planet on an orbit inclined with respect
  to a disc.  If the initial inclination of the orbit is larger than
  some critical value, the gravitational force exerted by the disc on
  the planet leads to a Kozai cycle in which the eccentricity of the
  orbit is pumped up to large values and oscillates with time in
  antiphase with the inclination.  On the other hand, both the
  inclination and the eccentricity are damped by the frictional force
  that the planet is subject to when it crosses the disc.  We show
  that, by maintaining either the inclination or the eccentricity at
  large values, the Kozai effect provides a way of delaying alignment
  with the disc and circularization of the orbit.  We find the
  critical value to be characteristically as small as about 
  20~degrees. Typically, Neptune or lower mass planets would remain
  on inclined and eccentric orbits over the disc lifetime, whereas
  orbits of Jupiter or higher mass planets would align and
  circularize.  This could play a significant role in planet formation
  scenarios.
\end{abstract}


%
%

\section{Introduction}
\label{sec:intro}

At the time of writing, more than 800 extrasolar planets have been
detected around main sequence stars.  Most of these objects have been
observed through radial velocity measurements, and about one third of
them have also been detected through transit measurements.  In
addition, the Kepler mission has so far released about 2300 planet
candidates (Borucki et al. 2011, Batalha et al. 2012).  Only a few of
these objects have been confirmed by radial velocity measurements so
far, but the number of false positive is expected to be very small.
The rate at which extrasolar planets are being discovered is
accordingly increasing very sharply.

The angle between the sky projections of the stellar spin axis and the
orbit normal (that we will call hereafter the {\em projected
  inclination angle}) has been measured using the Rossiter--McLaughlin
effect for 53 planets (Albrecht et al. 2012 and references therein).
About one third display significant misalignment with retrograde
orbits being indicated in some cases.
Misalignment, therefore, is common, at least for short period systems
for which measurements have been made.  Note however that the planets
in this sample are rather massive.  The lightest one has a mass of
about 25 earth masses, whereas
the other planets in the sample have masses ranging from a few tenths
of a Jupiter mass to several Jupiter masses.  By performing duration
ratio statistics on the Kepler planetary candidates, Fabrycky et
al. (2012) have concluded that pairs of planets in their sample are
aligned to within a few degrees.  As the Kepler catalog contains a
majority of super--earth and Neptune--like planets, it is possible
that misalignment is less common for lower mass planets.

According to the commonly accepted core accretion scenario, in which
planets form in a disc through the accretion of solid material into a
core followed, in the case of giant planets, by the capture of a
gaseous atmosphere, the orbit of planets should lie in the disc, and
therefore in the equatorial plane of the star.  In this model,
migration due to tidal interaction between disc and planets can be
invoked to explain the presence of giant planets on very short orbits.
However, such an interaction would not account for the occurrence of
any inclination of the orbit.  Therefore, this scenario is not likely
to apply to the hot Jupiters that are observed to be on inclined
orbits.  Planet formation through fragmentation of the disc might be
considered, but in that case the orbits are also expected to lie in
the plane of the disc.

One scenario that has been proposed for misaligning the orbits of hot
Jupiters that have formed in a disc lying in the stellar equatorial
plane relies on gravitational interaction with a distant stellar or
planetary companion that takes place after the disc has dissipated.
When the two orbits are not coplanar, the secular perturbation exerted
by the companion produces a Kozai cycle in which the eccentricity and
the inclination of the inner planet's orbit vary in antiphase.  If the
pericenter distance gets small enough, the orbit may become
circularized because of the tidal interaction with the central star,
and the orbit shrinks while, in some circumstances, keeping a high
inclination (Fabrycky \& Tremaine 2007, Wu et al. 2007, Naoz et
al. 2011).  In this model, a distant companion is needed which,
however, may not always be present in reality.  Other models rely on
planet--planet scattering (Chatterjee et al. 2008) or secular chaos
(Wu \& Lithwick 2011).

It has been proposed that the disc itself may be misaligned with
respect to the stellar equatorial plane.  That could happen if the
disc, at later times, were accreting material with angular momentum
misaligned with that of the star, as considered by Bate et al. (2010).
If at that stage there was enough mass in the disc to form planets,
their orbits would naturally be inclined with respect to the
stellar equatorial plane.  A similar scenario was studied by Thies et
al. (2011), who pointed out that close encounters of a disc accreting
from an extended envelope with another star could result in the disc
plane becoming tilted, possibly even to a retrograde orientation with
respect to its original one.  Planetary orbits inclined with respect
to the stellar equatorial plane would also result if the stellar spin
axis were tilted due to interaction with the disc (Lai et al. 2011,
Foucart \& Lai 2011).  Note however that, by comparing stellar
rotation axis inclination angles with the geometrically measured disc
inclinations for a sample of eight debri--discs, Watson et al. (2011)
have seen no evidence for a misalignment between the two.

Another possibility would be that the misaligned planets have formed
out of the disc through a fragmentation process occurring in the
protostellar envelope while it collapses onto the protostar, as
envisioned by Papaloizou \& Terquem (2001) and Terquem \& Papaloizou
(2002).  In this scenario, a population of planetary objects form
rapidly enough that their orbits can undergo dynamical relaxation on a
timescale on the order of a few $10^4$ years.  During the relaxation
process, most of the objects are ejected, while one to a few planets
become more bound to the central star.  Formation of hot Jupiters
through tidal circularization by the central star of a highly
eccentric orbit may occur (Papaloizou \& Terquem 2001).  The orbits of
the planets that are left in the system at the end of the relaxation
process display a range of eccentricities and inclinations (Adams \&
Laughlin 2003).  In this context, the misaligned planets interact with
the disc that has formed around the star during the protostellar
envelope collapse.  It is the subsequent dynamics of a system of this
type, consisting of a gaseous disc together with a planet with an
orbital plane misaligned with its midplane that we propose to
investigate here.  Note that we do not expect low mass planets to form
according to the scenario of Papaloizou \& Terquem (2001).  On the
other hand, in the scenario proposed by Bate et al. (2010), it may
happen that subsequent discs with different inclinations form and
dissipate, so that planets formed early in a disc may interact at
later time with a disc with different orientation.

In the present paper, we focus on a system with only one planet.
Multiple systems will be studied in future publications.  Some
preliminary work on the interaction of a planet on an inclined orbit
with a disc has been carried out by Terquem \& Ajmia (2010).  It was
found that the gravitational force exerted by the disc onto the planet
leads to a Kozai cycle in which the eccentricity and the inclination
of the orbit vary periodically with large amplitudes.  This indicates
that a planet's orbit which is inclined to start with may achieve high
eccentricity.  In their calculations, Terquem \& Ajmia (2010) adopted
a two dimensional flat disc model and ignored the frictional force
felt by the planet as it passes through the disc.  In the present
paper, we extend this work by modelling the disc in three dimensions
and including the frictional force.  Our goal is to understand under
what circumstances inclined orbits can be maintained when the disc is
present.  We consider planets with masses ranging from Neptune mass to
several Jupiter masses. The plan of the paper is as follows:
  
In section~\ref{sec:gravitation}, we describe the three dimensional
disc model used in the numerical simulations and the calculation of
the gravitational force exerted by the disc on a planet in an inclined
orbit.  A computationally convenient formulation, in terms of elliptic
integrals, is presented in an appendix.  We go on to give a brief
review of the Kozai mechanism in section~\ref{sec:Kozai}.  In
section~\ref{sec:friction}, we give an expression for the frictional
force exerted by the disc on the planet, and derive a damping
timescale based on a simplified analysis in
section~\ref{sec:analysis}.  In section~\ref{sec:simulations}, we
present the results of numerical simulations of the interaction
between a planet on an inclined orbit and a disc.  We first perform
simulations without the frictional force in section~\ref{sec:simgrav}
and compare the results to those of Terquem \& Ajmia (2010) that were
obtained for a two dimensional flat disc model.  In
section~\ref{sec:simfriction}, we include friction.  We show that,
when the planet's orbit starts with a low inclination (less than about
$23^{\circ}$) with respect to the disc, it becomes aligned with the
disc plane and circularized by the frictional force.  However, when
the inclination is initially high enough, the Kozai effect is present.
This pumps up the eccentricity of the planet's orbit maintaining
either the inclination or the eccentricity at large values.  As a
consequence, alignment of the orbital and disc planes, as well as
circularization of the orbit resulting from the frictional force, is
delayed.  In some cases, the orbit stays misaligned over the disc
lifetime.  More massive planets and planets further away from the star
align faster.  In addition, more massive discs favour alignment.  In
section~\ref{sec:discussion}, we discuss our results in the light of
the observations that have been reported so far.

%
%

\section{Interaction between a planet on an inclined orbit and a disc}
\label{sec:interaction}


\subsection{Gravitational potential and disc model}
\label{sec:gravitation}

We consider a planet of mass $\Mp$ orbiting around a star of mass
$\Ms$ which is itself surrounded by a disc of mass $\Md$.  The disc's
midplane is in the equatorial plane of the star whereas the orbit of
the planet is inclined with respect to this plane. We denote by $I$
the angle between the orbital plane and the disc's plane.
We suppose that the angular momentum of the disc is large compared to
that of the planet's orbit so that the effect of the planet on the
disc is negligible: the disc does not precess and its orientation is
invariable (see appendix~\ref{app:warp}).  We denote by $(x,y,z)$ the
Cartesian coordinate system centred on the star and $(r, \varphi, z)$
the associated cylindrical coordinates. The (axisymmetric) disc is in
the $(x,y)$--plane, its inner radius is $R_i$, its outer radius is
$R_o$ and its thickness (defining the region within which the mass is
confined) at radius $r$ is $2H(r)$.

The gravitational potential exerted by the disk at the location
$(r,z)$ of the planet is:

\be \Phi (r,z) = -G \int_{R_i}^{R_o} \int_{-H}^{H} \int_{0}^{2\pi}
\frac{\rho(r',z') r'dr'dz' d\phi'}{\sqrt{r^2+r'^2 - 2rr' \cos \phi'
    +(z-z')^2}},  
\label{potential}  \ee

\noindent where $G$ is the gravitational constant and $\rho$ is
the mass density in the disc.

We assume that $\rho$ falls off exponentially with $z^2$ near the
midplane with a cut off at $|z|=H(r),$ while decreasing as a power of
$r.$ Thus we adopt:

\be
\rho(r,z)=\left[ e^{\frac{1}{2}(1-z^2/H(r)^2)}-1 \right] \left(
  \frac{r}{R_o} \right)^{-n} \frac{\rho_0}{e^{1/2}-1},
\label{rho} 
\ee

\noindent where $\rho_0$ is a constant.  We have $\rho=0$ at the
surface of the disc, i.e. when $|z|=H$, and $\rho(r,0)=\rho_0
(r/R_o)^{-n}$ in the midplane.  This $z$--dependence of $\rho$ has
been chosen for simplicity, but we note that the details of how $\rho$
varies with $z$ do not matter.  What is important is the local disc
surface density at the location the planet passes through. This
largely determines the change in orbital energy resulting from the
dynamical drag force (as is apparent from equation (\ref{gammadyn}) of
section~\ref{sec:friction} below).

The mass density given above is discontinuous at $r=R_i$ and $r=R_o$,
as $\rho$ is zero outside the disc.  We have found that this could
introduce some numerical artefacts in the calculation of the disc's
gravitational force, so that in some runs we have replaced $\rho$ by
$\rho \times f(r)$ with:

\be
f(r) = \left[ 1 - \left( \frac{R_i}{r} \right)^{10} \right]
\left[ 1 - \left( \frac{r}{R_o} \right)^{20} \right].
\label{rhof}
\ee

\noindent The exponents 10 and 20 ensure that the edges are rather
sharp, so that we get quantitatively the same orbital evolution
whether the factor $f$ is used or not.

In the calculations presented below, we chose a value of the disc mass
$\Md$ and calculate $\rho_0$ using $\Md= \int \int \int_{disc} \rho
dV$. For the disc's semithickness, we chose a constant aspect ratio:
\be
H(r)=H_0 r ,
\label{H}
\ee
\noindent where $H_0$ is a constant.  The gravitational force per unit
mass exerted by the disc onto the planet is $ - \nab \Phi$.  In
appendix~\ref{app:potential}, we give convenient expressions for $ -
\nab \Phi $ in terms of elliptic integrals which can be readily
computed.

\subsection{The Kozai mechanism}
\label{sec:Kozai}

The Kozai effect (Kozai 1962, Lidov 1962) arises when an inner body on
an inclined orbit is perturbed by a distant companion. First derived
by Kozai to study the motion of inclined asteroids around the Sun
under perturbations from Jupiter, it has since found many applications
in astrophysics.  We are interested here in cases where the inner body
is a planet of mass $\Mp$. We denote $a$ its semimajor axis, $M'$ the
mass of the outer companion, assumed to be on a circular orbit, and
$a'$ its semimajor axis.   We consider the case $a' \gg a$.
The secular perturbation from the outer companion causes the
eccentricity $e$ of the inner planet and the mutual inclination $I$ of
the two orbits to oscillate in time in antiphase provided that the
initial inclination angle $I_0$ is larger than a critical angle $I_c$
given by:
\begin{equation} 
\cos^2 I_c = \frac{3}{5}.  
\label{Ic}
\end{equation}
\noindent The maximum value reached by the eccentricity is then given by:
\begin{equation}
e_{\rm max}=\left( 1- \frac{5}{3} \cos^2 I_0 \right)^{1/2},
\label{emax}
\end{equation}
\noindent and the time $t_{\rm evol}$ it takes to reach $e_{\rm max}$
starting from $e_0$ is (Innanen et al.~1997): 
\begin{equation}
\frac{t_{\rm evol}}{\tau} =  
0.42  
\left( \sin^2 I_0 - \frac{2}{5}  \right)^{-1/2} 
\ln \left( \frac{e_{\rm max}}{e_0}  \right),
\label{tevol}
\end{equation}

\noindent with the time $\tau$ defined  by:

\begin{equation}
  \tau=
  \left( \frac{a'}{a} \right)^3 \left( \frac{ M_{\star}}{ M'} \right) 
\frac{T}{2 \pi},
\label{tau}
\end{equation}

\noindent where $T$ is the orbital period of the inner planet.  If the
eccentricity oscillates between $e_{\rm min}$ and $e_{\rm max}$, then
the period of the oscillations $P_{\rm osc}$ is given by $P_{\rm osc}=
2t_{\rm evol}$ with $e_0 = e_{\rm min}$ in equation~(\ref{tevol}).

\noindent Since the two orbits are well separated, the component of
the angular momentum of the inner orbit perpendicular to the orbital
plane is constant.  As it is proportional to $\sqrt{1-e^2} \cos I$,
the oscillations of $I$ and $e$ are in antiphase.  Hereafter, we will
refer to this mechanism as the {\em classical Kozai effect}.

Terquem \& Ajmia (2010) found that the Kozai effect extends to the
case where the inner orbit is perturbed by the gravitational potential
of a disc, even when the orbit of the planet crosses the disc,
provided most of the disc mass is beyond the planet's orbit.  In that
case, $I$ is the angle between the orbital plane and the disc's
plane. They also showed that, in agreement with
equations~(\ref{tevol}) and~({\ref{tau}), the period of the
  oscillations decreased with $a.$ It was also found to decrease with
  increasing disc mass.  When the semimajor axis of the planet is
  small compared to the disc's inner radius, the evolution timescale
  is of the same form as that given by equation~(\ref{tevol}) but with
  \be \tau \propto \left( \frac{R_o}{a} \right)^3 \left( \frac{
      M_{\star}}{ \Md} \right) \frac{T}{2 \pi},
\label{taudisc}
\ee

\noindent where the coefficient of proportionality depends on the
functional form of $\rho$ and on the ratio $R_i/R_o$.

\subsection{Friction}
\label{sec:friction}

When the planet crosses the disc, as it has a relative velocity with
respect to the particles in the disc, it suffers a frictional force.
There are two types of drag acting on the planet: (i) an aerodynamic
drag, due to the fact that the planet has a finite size and suffers
direct collisions with the particles in the disc, and (ii) a dynamical
drag, due to the fact that particles in the disc are gravitationally
scattered by the planet.

Adopting a drag coefficient of unity, the aerodynamic drag force per
unit mass exerted on the planet located at $(r,\varphi,z)$, or
equivalently $(x,y,z)$, can be written as:

\be {\bf \Gamma}_{\rm aero} = - \frac{1}{2 M_p} \pi R_p^2 \rho(r,z)
v_{\rm rel} {\textbf v}_{\rm rel}, \ee
  
\noindent where $R_p$ is the planet's radius and ${\textbf v}_{\rm
  rel}$ is the relative velocity of the planet with respect to the
particles in the disc at the location $(x,y,z)$.  If we denote
${\textbf v}=(v_x,v_y,v_z)$ the velocity of the planet, then ${\textbf
  v}_{\rm rel} = ( v_x+y \Omega, v_y- x \Omega, v_z )$, where
$\Omega=\sqrt{G \Ms/r^3}$.

The dynamical drag is the gravitational force exerted by the particles
in the disc on the planet that is associated with the scattering and
change of location that occurs as a result of the passage of the
planet.  The main contribution to this force comes from the particles
located in the vicinity of the planet at the time it crosses the
disc. Note that the gravitational force from these particles has
already been included in $ - \nab \Phi $ (see
section~\ref{sec:gravitation}), but as we have ignored the motion of
the particles in the disc relative to the planet when calculating this
force it is conservative and does not capture the change of energy of
the planet's orbital motion.  The problem is similar to dynamical
friction in a collisionless medium (e.g., Binney \& Tremaine 1987).

When the inclination angle of the planet's orbit with respect to the
disc is not very small, $v_{\rm rel} \sim \sqrt{G \Ms / a}$, with $a$
being the semimajor axis of the planet's orbit,  is supersonic.
In this case, the dynamical friction force per unit mass acting on  the
planet can be written as (Ruderman \& Spiegel 1971, Rephaeli \&
Salpeter 1980, Ostriker 1999):

\be {\bf \Gamma}_{\rm dyn} = - 4 \pi G^2 M_p \rho(r,z)  \left| \ln
    \frac{H(r)}{R_p}  \right|
\frac{{\textbf v}_{\rm rel}}{v_{\rm rel}^3}.
\label{gammadyn}
\ee
The ratio of the two frictional forces is thus:
\be \frac{\Gamma_{\rm aero} }{\Gamma_{\rm dyn}} \sim \frac{1}{8 \left|
    \ln
    \frac{H(r)}{R_p} \right| } \left( \frac{R_p}{a} \right)^2 \left(
  \frac{\Ms}{\Mp} \right)^2 ,
\ee
\noindent where we have replaced $v_{\rm rel}^2$ by $G \Ms / a$.  In
this paper, we will focus on planets with masses at least that of
Neptune, for which the previous expression gives $ \Gamma_{\rm aero} /
\Gamma_{\rm dyn} \ll 1$ for the orbital parameters we consider.
Therefore, friction is dominated by the dynamical term even though a
large value of unity was adopted for the drag coefficient.

\subsection{Evolution timescale}
\label{sec:analysis}

The velocity $v$ of the planet is, in first approximation, the
Keplerian velocity around the star, i.e. $v \simeq \sqrt{G \Ms /a}$.
To simplify matters, we approximate the relative velocity by its
vertical component so that $v_{\rm rel} \sim v_z \simeq v \sin I
\simeq \sqrt{G \Ms /a} \sin I$.


When the planet's orbit has a significant eccentricity, the relative
velocity is in general larger as the planet crosses the disc closer to
pericenter than apocenter and its horizontal components (in the disc's
plane) are then important.  We define a damping timescale for the
planet in the absence of forces other than friction as:

\be \tau_{\rm damp} = \left( \frac{1}{v} \frac{dv}{dt} \right)^{-1} =
\frac{v}{ \Gamma_{\rm dyn}} .  \ee

\noindent This is the characteristic timescale on which the frictional
force ${\bf \Gamma}_{\rm dyn}$ damps the velocity of the planet.  Note
that $\tau_{\rm damp} = 2 a (da/dt)^{-1}$.  With the above
approximation for $v_{\rm rel}$  and $ \left| \ln(H/R_p) \right| \simeq 6$, 
which corresponds to giant planets at around 10~au in a disc with
$H(r) = 2.5 \times 10^{-2} r $, as we consider below in the numerical
calculations, we get:

\be \tau_{\rm damp} \sim \frac{\Ms ^2 \sin ^2 I}{24 \pi \Mp a^3
  \rho(r,z)} \frac{T}{2 \pi}, 
\label{taudamp1}
\ee

\noindent where $T$ is the orbital period of the planet.  

To proceed further, for the purpose of getting a convenient analytical
estimate of $\tau_{\rm damp}$, we approximate the vertical dependence
of the mass density by a $\delta$--function in $z$ and set $\rho(r,z)
=2 \delta (z)H(r) \rho_0 (r/R_o)^{-n}.$ Then we get $\Sigma(r) =
\int_{-H}^H \rho(r,z) dz =2 \rho_0 (r/R_o)^{-n} H(r)$.  Using
equation~(\ref{H}), we can then calculate the disc mass to be $ \Md =
\int_{R_i}^{R_o} \Sigma(r) 2 \pi r dr \simeq 8 \pi \rho_0 H_0 R_o^3/3,$
where we have used $n=3/2$, as in the numerical simulations below.
That allows us to express $\rho_0$ in term of $\Md$, such that $\rho_0
\simeq 3 \Md / (8 \pi H_0 R_o^3)$.  In accordance with equation
(\ref{rho}), we identify the midplane mass density as $\rho(r,0) =
\rho_0 (r/R_o)^{-n}\simeq 3 \Md (r/R_o)^{-n} / (8 \pi H_0 R_o^3)$. \\
We now suppose that the planet's orbit is not too eccentric so that
$\rho$ can be evaluated at $r=a$ in the expression of $\tau_{\rm
  damp}$.  Finally, equation~(\ref{taudamp1}) gives: \be \tau_{\rm
  damp} \sim \frac{ \Ms ^2}{ 9 M_p \Md} \left( \frac{R_o}{a}
\right)^{3/2} H_0 \sin ^2 I \frac{T}{2 \pi}.
\label{taudamp}
\ee This expression may not give the correct quantitative value of the
damping timescale, because of the approximations that have been used
in deriving it, but it gives the scaling of $\tau_{\rm damp}$ with the
different parameters.  Also, although the dependence on the
eccentricity has not been taken into account in this expression, as
pointed out above, we expect the damping timescale to be larger when
the eccentricity is larger.

\section{Numerical simulations}
\label{sec:simulations}

To study the evolution of the system (star, planet, disc), we use the
$N$--body code described in Papaloizou~\& Terquem~(2001) in which we
have added the gravitational and frictional forces exerted by the disc
on the planet.

The equation of motion for the planet is:
\begin{equation} {d^2 {\bf r} \over dt^2} = -{GM_\star{\bf r} \over
    |{\bf r}|^3} - \mbox{\boldmath $\nabla$} \Phi + {\bf \Gamma}_{\rm
    aero} + {\bf \Gamma}_{\rm dyn}  +
  {\bf \Gamma}_{t,r} - {GM_p{\bf r} \over |{\bf r}|^3} \; .
\label{emot}
\end{equation} 

\noindent 
The last term on the right--hand side is the acceleration of the
coordinate system based on the central star.  It arises because the
center of mass of the system does not coincide with that of the star.
This term takes into account only the force exterted by the planet
onto the star, and neglects the net force of the disc on the central
star, which would of course require a calculation of the disc response
to the planet's perturbation.  Tides raised by the star in the planet
and relativistic effects are included through ${\bf \Gamma}_{t,r}$,
but they are unimportant here as the planet does not approach the star
closely.  Equation~(\ref{emot}) is integrated using the
Bulirsch--Stoer method.  The integrals over $\phi$ involved in
$\mbox{\boldmath $\nabla$} \Phi$ are calculated using elliptic
integrals (see appendix~\ref{app:potential}). The integrals over $r$
and $z$ are calculated with the Romberg method (Press et al. 1993).

The planet is set on a circular orbit at the distance $r_p$ from the
star.  When the planet does not pass through the disc, e.g. when there
is no friction, the orbital energy is conserved and $r_p$ is equal to
the planet's semimajor axis $a$ throughout the evolution of the
system.  The initial inclination angle of the orbit with respect to
the disk is $I_0$.  In the simulations reported here, we have taken
$\Ms=1$~M$_{\odot}$, a radial power law with exponent $n=3/2$ for the
disc mass density $\rho$ (see eq.~[\ref{rho}]), a disc aspect ratio
$H_0=2.5 \times 10^{-2}$ and a disc outer radius $R_o=100$~au.

We will consider a planet with either (i) $M_p = 10^{-3}$~M$_{\odot}
\equiv 1 \; \MJ$ and $R_p=7 \times 10^9$~cm (Jupiter), (ii) $M_p=5
\times 10^{-5}$~M$_{\odot} \equiv 1 \; \MN$ and $R_p=2.5 \times
10^9$~cm (Neptune), (iii) $M_p=10^{-2}$~M$_{\odot}=10 \; \MJ $ and
$R_p=8.4 \times 10^9$~cm, i.e. 1.2 times Jupiter's radius.  These
latter values approximately correspond to the mass and radius of the
planet XO--3~b, which has been detected both in transit and using
radial velocity measurements (Johns--Krull et al. 2008, Winn et
al. 2008).

\subsection{Gravitation only}
\label{sec:simgrav}

To get a benchmark, we first consider the evolution of the planet's
orbit in the case where friction is ignored (i.e. $\Gamma_{\rm aero}$
and $\Gamma_{\rm dyn}$ are set to zero in eq.~[\ref{emot}]).  This
applies when the planet's distance to the star is at all times smaller
than the disc inner radius (e.g., planet orbiting in a cavity).  These
calculations are similar to those performed by Terquem \& Ajmia (2010)
and a comparison is made below.

Figure~\ref{fig1} shows the time evolution of the eccentricity $e$ and
inclination $I$ for $\Mp =1 \; \MJ$, $r_p = 5$~au, $\Md
=10^{-2}$~M$_{\odot}$, $R_i = 10$~au and $I_0 = 50^{\circ}$.

\begin{figure}
\centering \includegraphics[scale=0.5]{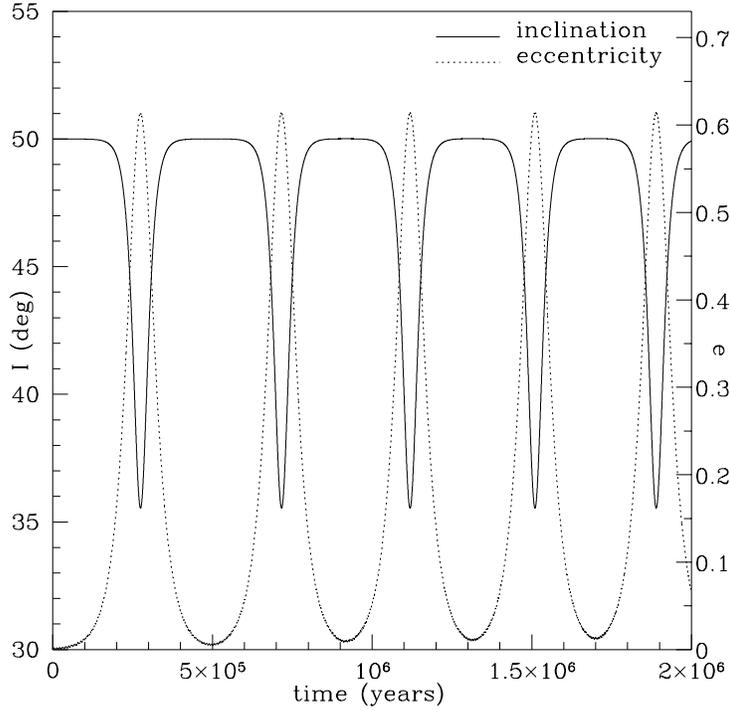}
\caption{Eccentricity $e$ (dotted line) and inclination angle $I$ (in
  degrees, solid line) versus time (in yr) in the absence of friction
  for $\Mp = 1 \; \MJ$, $r_p = 5$~au, $\Md
  =10^{-2}$~M$_{\odot}$, $R_i = 10$~au and $I_0 = 50^{\circ}$. }
\label{fig1}
\end{figure}

The inclination angle varies between $I_{\rm min}=I_c$ and $I_{\rm
  max}$, where $I_c$ is the critical value of $I_0$ below which
eccentricity growth is not observed.  The fact that $I_{\rm min}=I_c$
is illustrated in figure~\ref{fig2}, which shows the time evolution of
the inclination $I$ for $\Mp =1 \; \MJ$, $r_p = 7$~au, $\Md
=10^{-2}$~M$_{\odot}$, $R_i = 1$~au and $I_0$ varying
between $23^{\circ}$ and $50^{\circ}$.  For these parameters,
oscillations of the eccentricity and inclination disappear once $I_0$
becomes smaller than $23^{\circ}$, which means that $I_c=23^{\circ}$.
For values of $I_0$ well above $I_c$, the inclination is seen to
oscillate between $I_{\rm min}= I_c$ and $I_{\rm max}$.  When $I_0$ 
decreases however, the amplitude of the oscillations
is such that $I_{\rm min}$ gets a bit larger than $I_c$.  

\begin{figure}
\centering \includegraphics[scale=0.5]{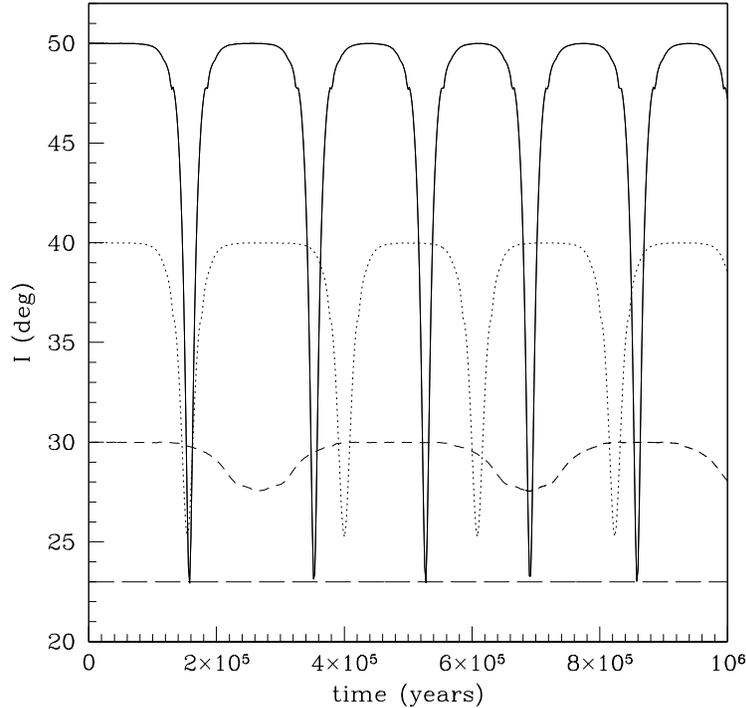}
\caption{Inclination $I$ (in degrees) versus time (in yr) in the
  absence of friction for $\Mp = 1 \; \MJ$, $r_p = 7$~au, $\Md
  =10^{-2}$~M$_{\odot}$, $R_i = 1$~au and $I_0 = 50^{\circ}$ (solid
  line), $ 40^{\circ}$ (dotted line), $ 30^{\circ}$ (short--dashed
  line) and $ 23^{\circ}$ (long--dashed line). For these parameters,
  $I_c=23^{\circ}$.  No oscillations are present for $I_0 < I_c$.}
\label{fig2}
\end{figure}

As already noted by Terquem \& Ajmia (2010), since $I_c$ depends on
the gravitational force exerted onto the planet, it varies with the
initial position of the planet $r_p$.  
When $r_p \ll R_i$, the conditions are similar to the classical Kozai
effect and the gravitational force from the disc can be approximated
by a quadrupole term.  In this case, it can be shown that
$I_c=39^{\circ}$ (eq.~[\ref{Ic}]).  When $r_p$ is larger though, the
quadrupole approximation is not valid anymore and $I_c$ gets smaller.
This is illustrated in figure~\ref{fig3}, which shows the critical
angle $I_c$ as a function of $r_p$, which here is the same as the
planet's semimajor axis since the orbit is initially circular and
there is no energy dissipation.  These calculations are done for $\Mp
=1 \; \MJ$, $\Md = 10^{-2}$~M$_{\odot}$, $R_i = 10$~au and
$r_p$ varying between $1$ and $35$~au.  For these parameters, the
Kozai effect disappears when $r_p$ becomes larger than $\sim$40~au, as
most of the mass is then no longer outside the orbit of the planet
(Terquem \& Ajmia 2010).

\begin{figure}
\centering \includegraphics[scale=0.5]{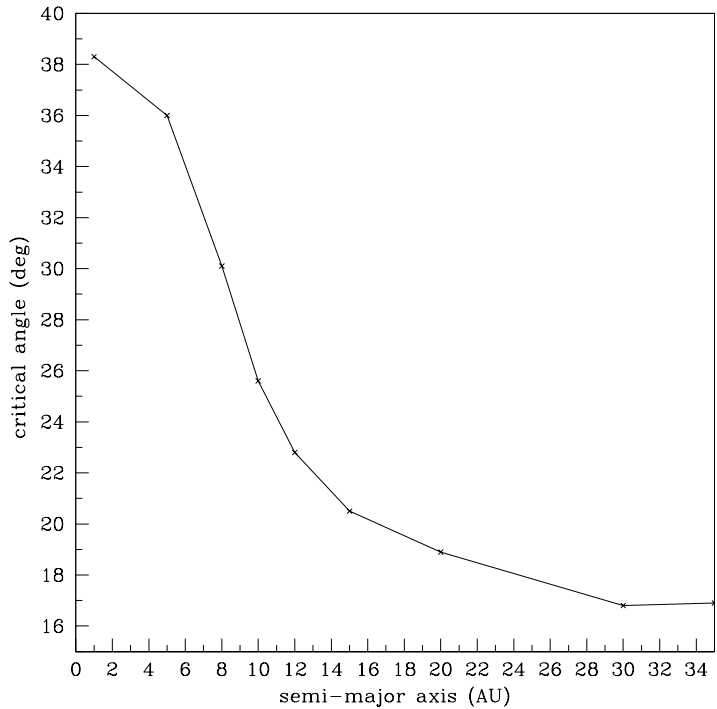}
\caption{$I_c$ (in degrees) versus $r_p$ (in au) in the absence of
  friction for $\Mp =1 \; \MJ$, $\Md =10^{-2}$~M$_{\odot}$,
  $R_i = 10$~au and $r_p$ varying between $1$ and $35$~au.  For $r_p
  \ll R_i$~au we recover the classical Kozai value $I_c=39^{\circ}$. }
\label{fig3}
\end{figure}

For comparison, we have also run the case displayed in
figure~\ref{fig1} by modelling the disc in two dimensions only, as in
Terquem \& Ajmia (2010).  The gravitational potential is then
calculated as arising from a distribution of mass with surface density
$\Sigma_{\rm 2D}(r) \propto r^{-p}$.  We take $p=1/2$  as in
the 3D calculations we have $\Sigma(r) \sim \rho(r)H(r) \propto
r^{-1/2}$.  There is very good agreement between the two sets of
calculations.  The extrema of $I$ and $e$ and the period of the
oscillations are roughly the same.

In the 3D case, we have also checked that the vertical structure of
the disc does not affect the oscillations.  Indeed, when the disc mass
is kept constant but $H_0$ is varied in equation~(\ref{H}), the
oscillations are unchanged.

\subsection{The effect of friction}
\label{sec:simfriction}

We are now going to study the effect of friction on the evolution of
the planet's orbit by taking into account $\Gamma_{\rm aero}$
and $\Gamma_{\rm dyn}$ in equation~(\ref{emot}).

Figure~\ref{fig4} shows the time evolution of the orbital parameters
$I$, $e$, semimajor axis $a$ and distance to pericenter $a(1-e)$ for
$\Mp =1 \; \MN$, $r_p = 7$~au, $\Md =10^{-2}$~M$_{\odot}$, $R_i =
1$~au and for two different values of $I_o$.  For these parameters,
$I_c=23^{\circ}$ (the critical angle does not depend on the mass of
the planet an therefore is the same as in fig.~\ref{fig2}).  As
discussed in section~\ref{sec:analysis}, the damping timescale
increases with $I$ and $e$.  Therefore, when $I_o>I_c$, as at all
times one of these parameters has a large value because of the Kozai
cycle, the damping timescale is longer than in the case $I_o<I_c$ and
$e=0$.  This is illustrated in figure~\ref{fig4}, where we see that
$a$ decreases much more rapidly when $I_o=20^{\circ}$ than when
$I_o=50^{\circ}$.  
In the case where there is no Kozai cycle,
the orbit stays circular and $I$ is damped faster as it becomes
smaller, so that the orbit aligns with the disc on a very short
timescale.  When $I_o>I_c$, Kozai oscillations are present as
expected, but the oscillations are damped because of friction.  We
observe that damping is much less efficient than in the
$I_o=20^{\circ}$ case, even though the inclination $I$ does reach
values that are not much larger than $20^{\circ}$ and for which the
damping timescale would be similar if the orbit were circular.  When
the Kozai cycle is present though, the eccentricity of the orbit is
large when $I$ is minimum, which results in the damping timescale
being maintained at large values.  Contrary to what happens when
$I_o<I_c$ then, the damping timescale stays roughly constant.  This is
illustrated in the plot of figure~\ref{fig4} that shows $a$ versus
time.  Ultimately, if the disc were present long enough, the orbit of
the planet would align with the disc and would be circularized ($I$
and $e$ vanish).  For a Neptune mass planet and the parameters used
here though, we find that misalignment can be maintained over the the
disc lifetime, which is of a few million years.  A Jupiter mass planet
would align faster, as discussed below.

These calculations show that, by pumping up the eccentricity of the
planet's orbit and maintaining either $I$ or $e$ at large values, the
Kozai effect provides a way of delaying alignment with the disc and
circularization of the orbit.

We see in figure~\ref{fig4} that the period of the oscillations
decreases with time.  If there were no dissipation, this period would
stay constant.  In the classical Kozai cycle, when the orbit of the
inner planet has a semimajor axis $a$, the time it takes to reach
$e_{\rm max}$ starting from $e_{\rm min}$, is proportional to $a^{-3/2}
\ln \left( e_{\rm max} /e_{\rm min} \right)$ (see eq.~[\ref{tevol}]
and~[{\ref{tau}]).
Although this formula has been derived for a quadrupolar gravitational
potential, it may be expected to give a general trend in the disc case
as well.  Friction reduces $a$, which tends to increase this
timescale.  But the amplitude of the cycle also decreases ($e_{\rm
  min}$ increases and $e_{\rm max}$ decreases), so that the net effect
is a shorter timescale.

As pointed out in section~\ref{sec:friction}, the expression for
dynamical friction given by equation~(\ref{gammadyn}) is valid only
when the relative velocity of the planet with respect to the particles
in the disc is supersonic.  This is not the case when $I$ is close to
zero and, therefore, the part of the curves in figure~\ref{fig4}
corresponding to $I$ close to zero can be interpreted only
qualitatively.

\begin{figure}
\centering \includegraphics[scale=0.7]{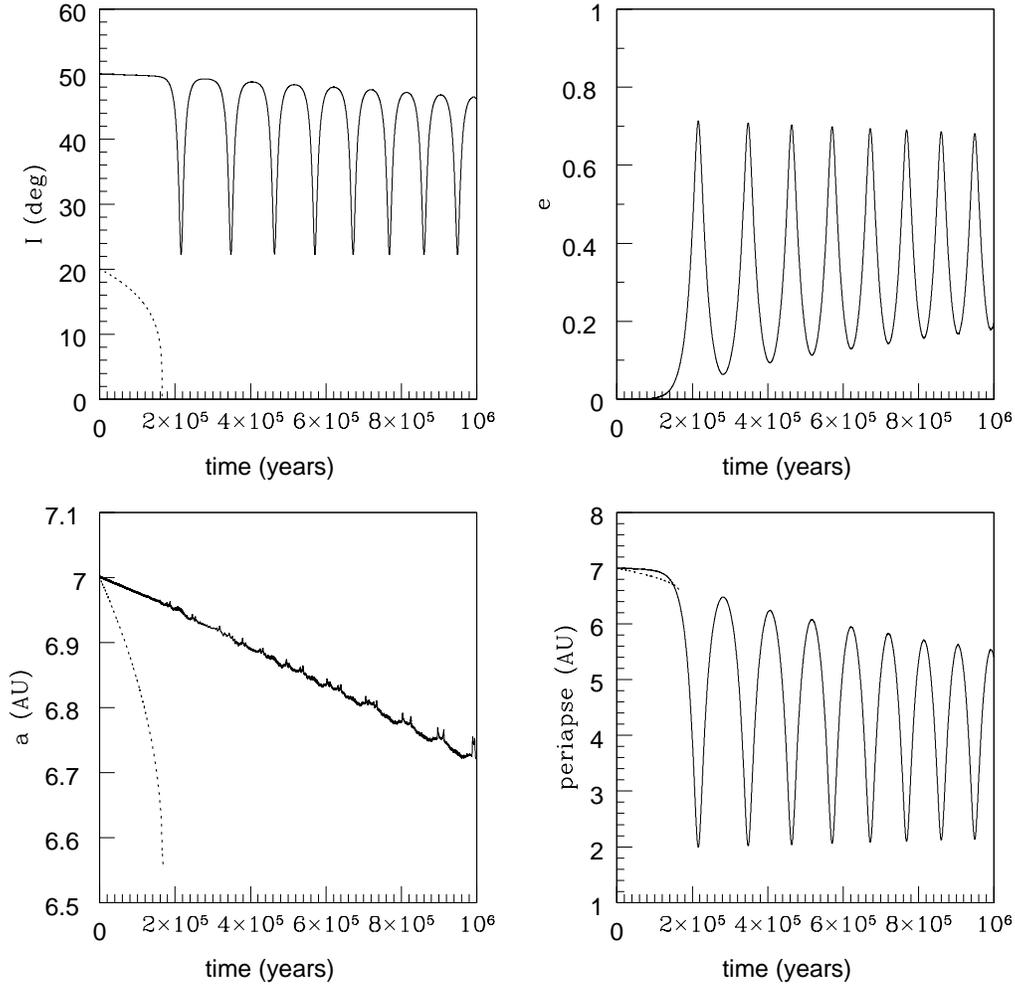}
\caption{Inclination $I$ (in degrees, upper left plot), eccentricity
  $e$ (upper right plot), semimajor axis $a$ (in au, lower left plot)
  and pericenter distance $a(1-e)$ (in au, lower right plot) versus
  time (in yr) for $\Mp =1 \; \MN$, $r_p=7$~au, $\Md
  =10^{-2}$~M$_{\odot}$, $R_i = 1$~au and $I_0=50^{\circ}$ (solid
  line) and $20^{\circ}$ (dotted line). For these parameters,
  $I_c=23^{\circ}$.  When $I_0=20^{\circ}$, $e=0$ throughout the
  evolution so the curve does not show on the plot.  The Kozai effect
  provides a way of pumping up $e$ and maintaining either $e$ or $I$
  at large values, and therefore delays circularization and alignment
  with the disc of the orbit.  }
\label{fig4}
\end{figure}

\subsubsection{Influence of the planet's mass:}

We now study the influence of the planet's mass on the evolution of
the orbit.  Figure~\ref{fig5} shows the time evolution of the
inclination $I$, eccentricity $e$ and semimajor axis $a$ for the same
parameters as in figure~\ref{fig4}, i.e.  $r_p = 7$~au, $\Md
=10^{-2}$~M$_{\odot}$, $R_i = 1$~au, $I_o=50^{\circ}$ and three
different values of $\Mp$.  As $I_c=23^{\circ}$ for these parameters,
we are in the regime where Kozai cycles are present.  As shown by
equation~(\ref{taudamp}), the damping timescale $\tau_{\rm damp}$ is
proportional to $1/M_p$.  The period $P_{\rm osc}$ of the Kozai cycle,
on the other hand, does not depend on the planet's mass (see
eq.~[\ref{tau}] and~[\ref{taudisc}]).  We therefore expect friction to
be very efficient for high mass planets for which the damping
timescale $\tau_{\rm damp}$ would be smaller than $P_{\rm osc}$, and
much less efficient for low mass planets for which $\tau_{\rm damp}
\gg P_{\rm osc}$.  This is borne out by the results displayed in
figure~\ref{fig5}.  Friction dominates the evolution for the planet
with $\Mp=10 \; \MJ$, for which $\tau_{\rm damp}$ is smaller than
$P_{\rm osc}$.  For a Jupiter mass planet, the two timescales are
comparable and the orbit aligns after a few oscillations, i.e. after a
few $10^5$ years.  For a Neptune mass planet, $\tau_{\rm damp} \gg
P_{\rm osc}$ and $I$ decreases very slowly.  In that case, the orbit
would stay misaligned over the disc lifetime.

As stated above, $P_{\rm osc}$ does not depend on the planet's mass.
However, we see from figure~\ref{fig5} that the oscillations for
$M_p=1 \; \MJ$ and $M_p= 1 \; \MN$ do not have the same
period.  This is because, as noted above, $P_{\rm osc}$ decreases as
damping reduces the amplitude of the oscillations.

The fact that dynamical drag leads to faster alignment of the orbits
for more massive planets was seen in the numerical simulations of Rein
(2012) who considered planets on highly inclined orbits but without
taking into account the gravitational interaction with the disc.

\begin{figure}
\begin{center}
\includegraphics[scale=0.7]{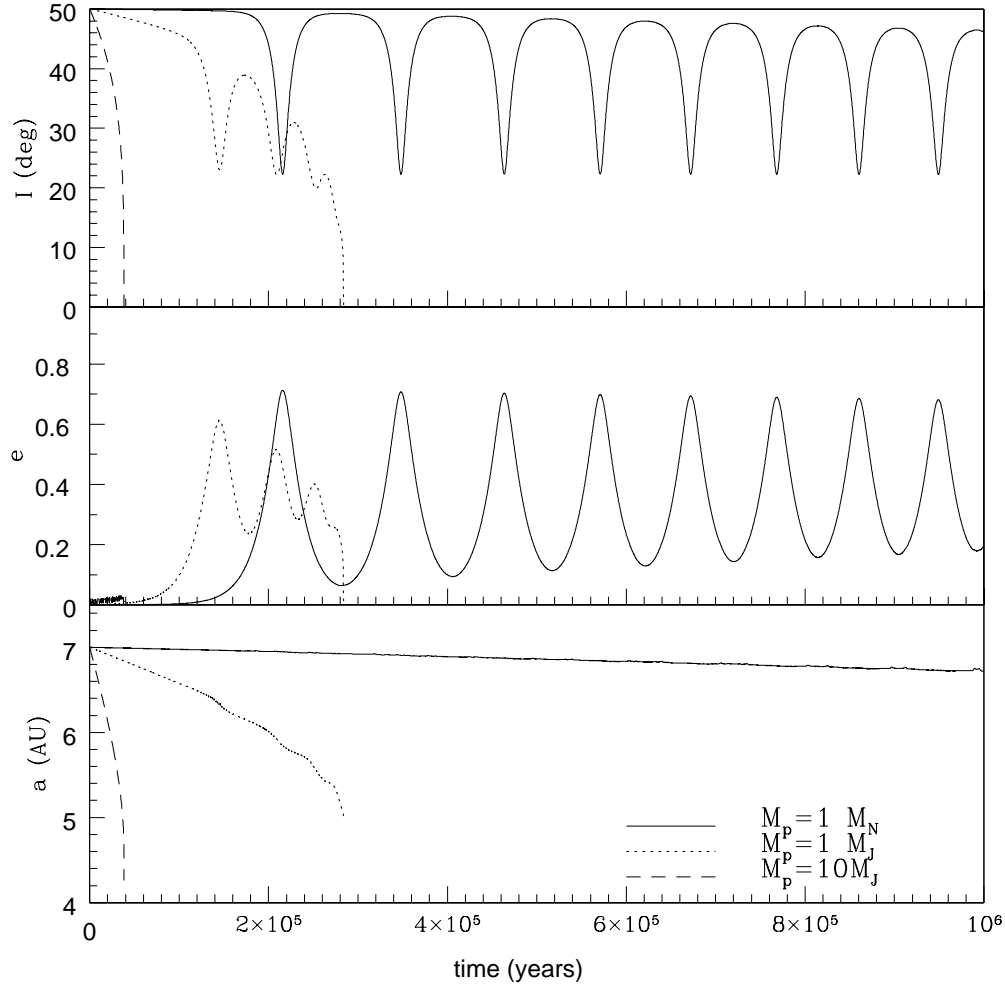}
\end{center}
\caption{Inclination $I$ (in degrees, upper plot), eccentricity
  (middle plot) and semimajor axis $a$ (in au, lower plot) versus time
  (in yr) for $r_p=7$~au, $\Md =10^{-2}$~M$_{\odot}$, $R_i = 1$~au,
  $I_0=50^{\circ}$ (same parameters as in fig.~\ref{fig4}) and $\Mp=1
  \; \MN$ (solid line), $1 \; \MJ$ (dotted line) and $10 \; \MJ$
  (dashed line). For these parameters, $I_c=23^{\circ}<I_0$ so that we
  are in the regime where Kozai cycles are present.  Friction damps
  the oscillations more efficiently for larger mass planets. Alignment
  of the orbit with the disc may not happen over the disc lifetime for
  smaller mass planets.}
\label{fig5}
\end{figure}


\subsubsection{Influence of the disc's mass:}

Figure~\ref{fig6} shows the time evolution of the inclination $I$ and
semimajor axis $a$ for $\Mp =1 \; \MN$, $r_p = 7$~au, $R_i
= 1$~au, $I_0=50^{\circ}$ (as in fig.~\ref{fig4}) and three different
values of $\Md$ ranging from $10^{-2}$ to 0.1~M$_{\odot}$.  Terquem \&
Ajmia (2010) showed that $I_c$ does not depend on $\Md$ when the other
parameters are kept fixed, so that we have $I_c=23^{\circ}$ here.
Since $I_0=50^{\circ}$, we are therefore in the regime of Kozai
cycles.  From equation~(\ref{taudamp}), we see that $\tau_{\rm damp}
\propto \Md^{-1}$.  On the other hand, we also have $P_{\rm osc}
\propto \Md^{-1}$, as expected from the classical Kozai effect and
verified in the disc's case by Terquem \& Ajmia (2010).  It follows
that alignment of the orbit with the disc, when it occurs, should
happen after the same number of oscillations whatever the disc's mass.
This is confirmed by the curves in figure~\ref{fig6} which are all a
time--scaled version of each other, with the more massive disc showing
the shortest evolution timescale.  It is therefore easier to keep a
planet on an inclined orbit in the presence of a less massive disc.

\begin{figure}
\begin{center}
\includegraphics[scale=0.7]{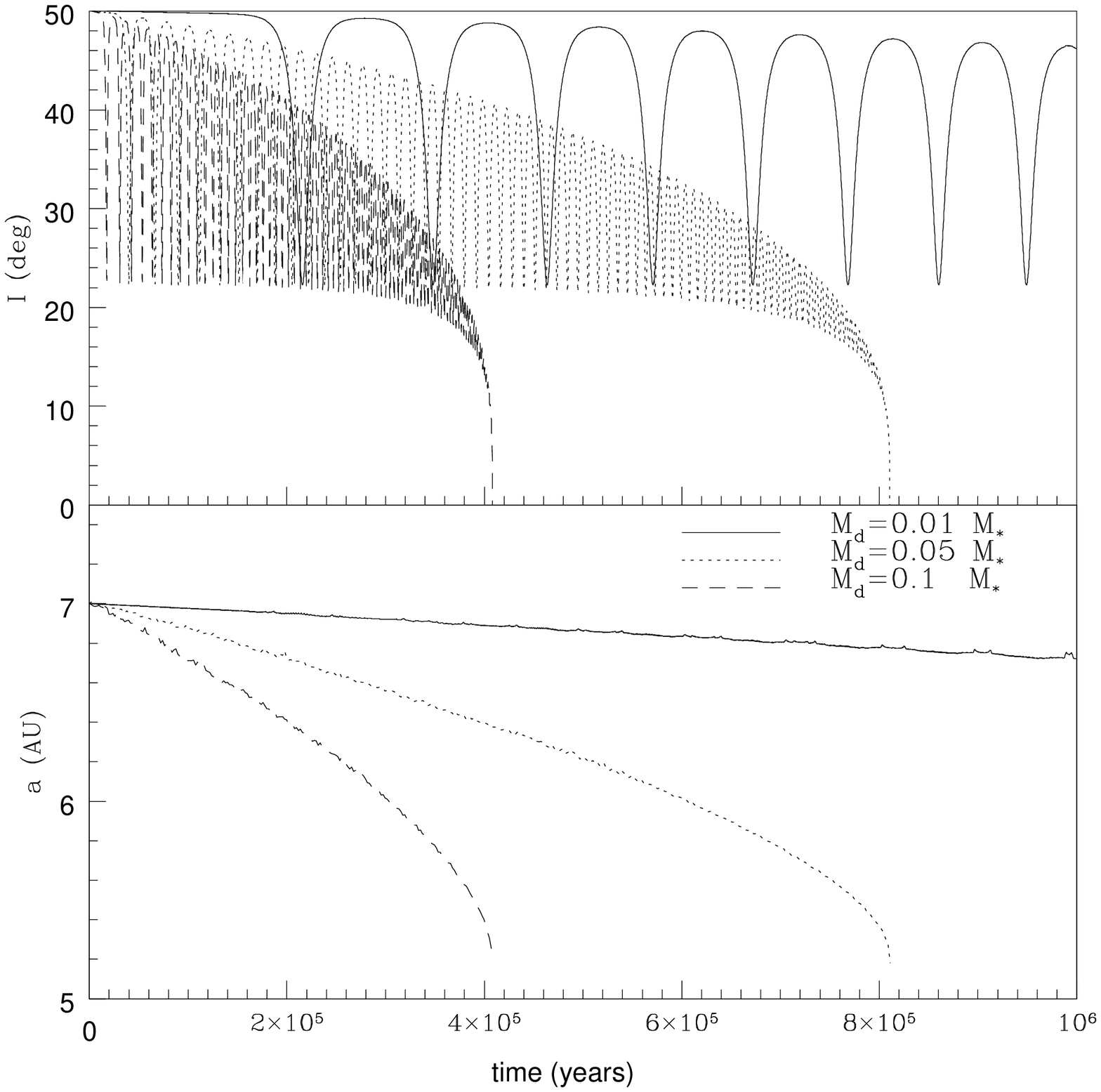}
\end{center}
\caption{Inclination $I$ (in degrees, upper plot) and semimajor axis
  $a$ (in au, lower plot) versus time (in yr) for $\Mp
  =1 \; \MN$, $r_p=7$~au, $R_i = 1$~au, $I_0=50^{\circ}$ (same parameters as
  in fig.~\ref{fig4}) and $\Md=0.1$~M$_{\odot}$
  (dashed line), $5 \times 10^{-2}$~M$_{\odot}$ (dotted line) and
  $10^{-2}$~M$_{\odot}$ (solid line).  For these parameters,
  $I_c=23^{\circ}<I_0$ so that we are in the regime where Kozai cycles
  are present. It is easier to keep a planet on an inclined orbit in
  the presence of a less massive disc.}
\label{fig6}
\end{figure}


\subsubsection{Influence of the planet's initial semimajor axis:}

Finally, we study the effect of the initial semimajor axis on the
orbital evolution.  Figure~\ref{fig7} shows the time evolution of the
inclination $I$ for $\Mp =1 \; \MJ$, $\Md=10^{-2}$~M$_{\odot}$, $R_i =
10$~au, $I_0=60^{\circ}$ and four different values of $r_p$ ranging
from 5 to 20~au.  For these parameters, we are in the regime of Kozai
cycles.  For $r_p=5$~au, the planet stays in the disc's inner cavity
so that it never experiences frictional forces when it goes through
the disc's plane.  The orbital evolution is therefore that of a Kozai
cycle with no friction and a minimum angle of $37^{\circ}$, consistent
with the results of figure~\ref{fig3}.  For $r_p=8$~au, although the
distance to the star does get larger than $R_i$ when $e>0.25$, as it
happens the planet is always in the disc's inner cavity when it
crosses the disc's plane, so that here again friction does not play a
role.  Like in the previous case, the minimum angle for this cycle is
consistent with the results of figure~\ref{fig3}.  As noted above,
$P_{\rm osc} \propto a^{-3/2}$ in the classical Kozai case.  We verify
here that $P_{\rm osc}$ does indeed decrease when $a$ increases (in
agreement with Terquem \& Ajmia 2010).  For $r_p=12$ and 20~au, the
planet crosses the disc so that the oscillations are damped by the
frictional force.  As shown by equation~(\ref{taudamp}), the damping
timescale $\tau_{\rm damp}$ does not depend on $a$.  This is
consistent with the curves displayed in figure~\ref{fig7} at early
times, before the eccentricity of the orbits grow, as they show that
the damping timescale is indeed the same for $r_p=12$ and 20~au.  But
as the inclination gets smaller for the larger value of $r_p$ ($I_{\rm
  min}$ is smaller and the planet spends more time in the disc),
$\tau_{\rm damp}$ gets also shorter for this value of $r_p$.
Alignment of the orbit with the disc is then faster at larger
distances from the star.

\begin{figure}
\centering \includegraphics[scale=0.5]{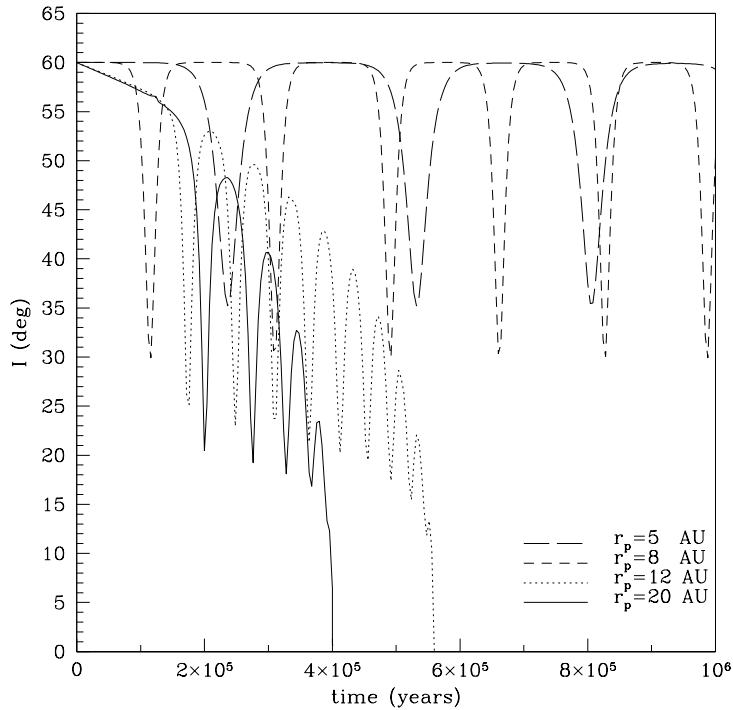}
\caption{Inclination $I$ (in degrees) versus time (in yr) for $\Mp
  =1 \; \MJ$, $\Md=10^{-2}$~M$_{\odot}$, $R_i = 10$~au,
  $I_0=60^{\circ}$ and $r_p=20$~au (solid line), 12~au (dotted line),
  8~au (short dashed line) and 5~au (long dashed line). For these
  parameters, we are in the regime of Kozai cycles. Alignment of the
  orbit with the disc is faster at larger distances from the star.}
\label{fig7}
\end{figure}

%
%

\section{Discussion}
\label{sec:discussion}

  We have investigated the dynamics of a planet on an orbit inclined
  with respect to a disc.  If the initial inclination of the orbit is
  larger than some critical value, the gravitational force exerted by
  the disc on the planet leads to a Kozai cycle in which the
  eccentricity of the orbit is pumped up to large values and oscillates
  with time in  antiphase  with the inclination.  On the other
  hand, when the planet goes through the disc, it suffers a frictional
  force that results in a loss of orbital energy.  As a consequence,
  the inclination and the eccentricity of the orbit are damped and the
  semimajor axis decreases.  The goal of this paper was to study on
  what timescale orbits inclined with respect to a disc would align
  with it.

  The calculations presented in this paper show that, by pumping up
  the eccentricity of the planet's orbit and maintaining either $I$ or
  $e$ at large values, planets in orbits undergoing Kozai cycles
  maintain large velocities relative to the disc as they pass
  through it, so delaying alignment with the disc and circularization
  of the orbit.

  For the parameters used in this paper, which are typical of
  protostellar discs, it was found that Neptune mass planets would
  remain on inclined orbits over the disc lifetime.  Jupiter mass
  planets, however, tend to align faster, as the damping timescale is
  shorter for more massive planets.  Note however that we have not
  taken into account the fact that the disc dissipates progressively
  over time.  As damping is less efficient in less massive discs,
  Jupiter mass planets could remain misaligned if the disc's mass were
  decreasing sufficiently fast.  We have also found that alignment of
  the orbits was faster at larger distances from the star.  

  So far, the only planets that have been found on inclined orbits are
  rather massive (with a mass $\sim \; \MJ$) and on short period
  orbits.  This of course is a result of observational bias, as the
  inclination is measured so far only for transiting planets.  As
  the results of this paper suggest, if these planets had been on
  inclined orbits when the disc was present, they probably would have
  aligned if they had crossed the disc.  Therefore, either (i) the
  orbits became inclined after the disc had dissipated, (ii) the disc
  in which the planets formed was misaligned with the stellar
  equatorial plane, or (iii) the planets formed on inclined orbits
  with short enough periods that they crossed the disc's plane only in
  an inner cavity.  Jupiter mass planets formed on inclined orbits at
  large distances from the star would be expected to have aligned with
  the disc unless the formation took place near the end of the life of
  the disc.

The planets in the Kepler sample seem to have both low inclinations
(Fabrycky et al. 2012) and low eccentricities (Kane et al. 2012).
Radial velocity surveys also show lower eccentricities for lower mass
planets.  According to the results of our paper, this strongly
suggests that these planets, with masses mainly at most $\sim \; \MN$,
have formed and stayed in the original disc.  Had they been in a
sufficiently inclined orbit at some point, this would have recurred
over the disc lifetime, and the orbit would have had episodes of high
eccentricity.  Thus a population in both highly eccentric and inclined
orbits would be expected.  Measurement of inclination angles for
longer period orbits will enable the evaluation of proposed aspects of
planet formation scenarios.

Finally, we remark that in the present paper we have considered only
one planet interacting with the disc.  In a subsequent paper, we will
investigate the dynamics of multiple systems.

%
%


%

\appendix

\section{Response of the disc to the gravitational interaction with
  the planet}
\label{app:warp}

We here estimate the warping response of a disc of the type we
consider to a planet in an inclined  circular orbit. The disc is
assumed to contain significantly more angular momentum than the planet
and obey a barotropic equation of state.  We show that provided the
inverse of the orbital precession frequency is less than the local
disc sound crossing time and the mass of the planet is less than the
disc mass in its neighbourhood, the range of inclinations excited is
expected to be small.

 \subsection{Governing equations}

 The basic governing equations are the equations of continuity and
 motion for a barotropic gas in the form
\begin{eqnarray}
  \frac{\partial {\rho} }{\partial t}+\nabla\cdot \rho{\bf v}&=&0,\label{A0} \\
  \frac{\partial {\bf v} }{\partial t}+{\bf v}\cdot\nabla{\bf v}&=&-\frac{1}{\rho}\nabla{P}-\nabla{\Phi}-\nabla{\Phi_*},\label{Basiceqs}
\end{eqnarray}
where $\Phi_*= -GM_*/|{\bf r}|$ is the gravitational potential due to
the central star and \be \Phi= -\frac {GM_p}{\sqrt{r^2+R^2
    -2rR\cos(\phi -\phi_p)+(z-z_p)^2}} \ee is the potential due to the
planet which is treated as a perturbation for which we calculate the
linear response below (the indirect term does not contribute to the
warping and so may be dropped).  Here the cylindrical coordinates of
the planet are $(R,\phi_p, z_p).$

We adopt a Fourier decomposition in azimuth and time of the form \be
\Phi =\sum_{m >0} \Phi_{m} \exp [ {\rm i}(m\phi +\omega_{p,m}t) ] +
cc, \label{Fourier} \ee where $+ cc$ indicates the addition of the
complex conjugate, $m$ is the azimuthal mode number and $\omega_{p,m}$
is an associated frequency corresponding to a pattern speed
$-\omega_{p,m}/m.$ For global warps we are interested in $m=1$ and
pattern speeds that correspond to a slow precession of the planetary
orbit.  The precession period of the line of nodes, which for the
purposes of this section is assumed to be given, is related and
comparable to the period of Kozai oscillations when these occur.  We
adopt the time or orbit average of the coefficient $\Phi_{1}$ which is
appropriate for a discussion of the secular evolution of global warps.
We perform a corresponding Fourier decomposition for the response
perturbations to the disc which are taken to have a $\phi$ and $t$
dependence through a multiplicative factor $\exp [ {\rm i}(m\phi
+\omega_{p,m}t) ].$ From now on this factor will be taken as read and
we drop the subscript $m$ on quantities as this is taken to be unity.

\subsection{The  disc inclination response}

Denoting perturbations with a prime, linearization of the equations of
motion (\ref{Basiceqs}) for the response to the perturbing potential,
$\Phi,$ gives
\begin{eqnarray}
  {\rm i}(\omega_p+\Omega)v_r' -2\Omega v_{\phi}' &= &
- \frac{\partial W }{\partial r}\nonumber \\
  {\rm i}(\omega_p+\Omega)v'_{\phi} +\frac{\kappa^2}{2\Omega} v_{r}' &= &
- \frac{{\rm i} W }{ r}\nonumber \\
  {\rm i}(\omega_p+\Omega)v_z'&= &- \frac{\partial W }{\partial
    z} \label{lineom}
\end{eqnarray}
where $W=P'/\rho +\Phi$ and $\kappa$ is the epicyclic frequency.
Solving for the velocity perturbations, we obtain
\begin{eqnarray}
  v_r'  &= &-{\rm i}\frac{(\omega_p+\Omega) 
\partial W /\partial r+ 2\Omega W / r} 
  { \kappa^2-(\omega_p+\Omega)^2}  \nonumber \\
  v'_{\phi}  &= &\frac{(\kappa^2/(2\Omega ))\partial W /\partial r+  
(\omega_p+\Omega)W / r} 
  {\kappa^2-(\omega_p+\Omega)^2 } \label{lineom1}
\end{eqnarray}
\begin{figure}
\centering \includegraphics[scale=0.7]{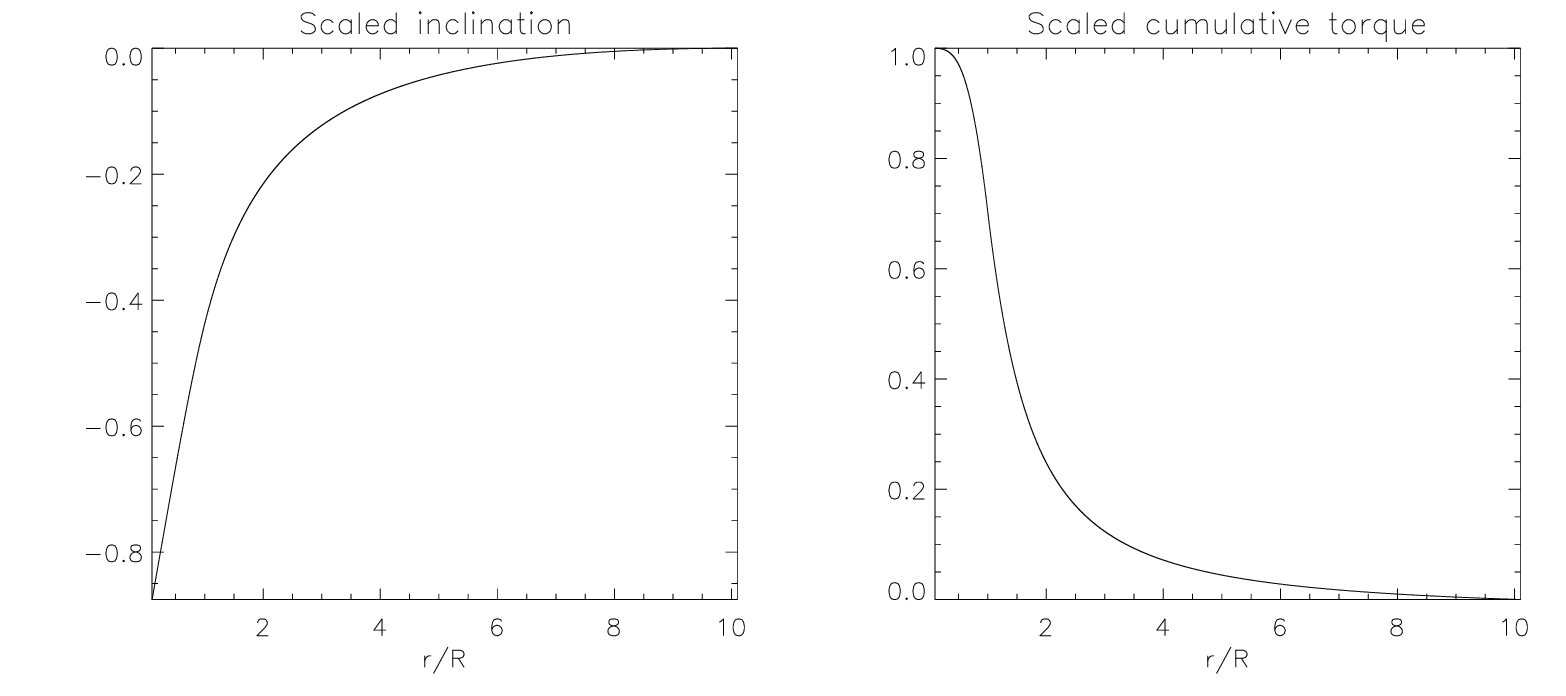}
\caption{ The left hand panel shows the inclination, normalized by the
  factor $10^3(M_p/M_*)(0.05R/H(R))^2 \omega_p R^{3/2}/\sqrt{GM_*},$
  as a function of $r/R$ for the calculation described in the text.
  This was calculated under the condition that it tended to zero as
  $r\rightarrow \infty.$ The right hand panel shows the function
  $T(r/R)/T(0)$ which represents the magnitude of the torque due to
  the planet acting on the disc in the radial interval $(r,\infty)$
  normalized by its value as $r\rightarrow 0.$ For both panels the
  inclination of the circular planetary orbit of radius $R$ to the
  plane of the disc was $\pi/4.$ }
\label{figA1}
\end{figure}
As we are interested in a disc that is close to Keplerian rotation,
with $\omega_p \ll \Omega,$ and $\kappa \sim \Omega,$ we neglect
$\omega_p$ and set $\kappa =\Omega$ in the numerators above, and
replace $\kappa^2-(\omega_p+\Omega)^2$ by
$2\Omega(\kappa-\omega_p-\Omega)$ in the denominators to obtain
\begin{eqnarray}
  v_r'  &= &-{\rm i}\frac{\partial W /\partial r+ 2W / r} 
  {2\left (\kappa-\omega_p- \Omega \right) }\nonumber \\
  v'_{\phi}  &= &\frac{\partial W /\partial r+  2W / r} 
  {4\left (\kappa-\omega_p-\Omega \right) } \label{lineom1a}
\end{eqnarray}
Following our previous work (eg Papaloizou \& Terquem 1995, Larwood et
al. 1996, Nelson \& Papaloizou 1999) we seek a solution for which
$v_z'$ is independent of $z$ to within a correction of order
$(H/r)^2.$ Then to within the same order of accuracy we may integrate
the linearized $z$ component of the equation of motion to give \be W
=-{\rm i}(\omega_p+\Omega) z v_z'.\label{Wlin} \ee We now write down
the linearized continuity equation in the form \be \frac{{\rm
    i}(\omega_p+\Omega)\rho (W - \Phi)}{c_s^2} = -\nabla\cdot(\rho
{\bf v}'), \label{adpert} \ee where we have used $P'= c_s^2\rho',$
with $c_s^2=dP/d\rho.$ Multiplying (\ref{adpert}) by $z$ and
integrating over the vertical extent of the disc, we get

\begin{equation}
\int^{\infty}_{-\infty}\frac{{\rm i}(\omega_p+\Omega) \Phi \rho z}{c_s^2} dz
  =\int^{\infty}_{-\infty}\frac{(\omega_p+\Omega)^2 v_z' \rho z^2}{c_s^2} dz   -
\int^{\infty}_{-\infty} v_z'\rho dz + \hspace{2mm} \nabla_{\perp}\cdot 
\left(\int^{\infty}_{-\infty}z{\bf v}'_{\perp}\rho dz\right),
\label{vertave}
\end{equation}
where the perpendicular velocity perturbation is ${\bf
  v}'_{\perp}=(v_r',v_{\phi}',0).$
Making use of equations (\ref{lineom1a}), (\ref{Wlin}) with $\omega_p$
neglected in the latter, and vertical hydrostatic equilibrium for the unperturbed
state, we obtain
\begin{equation}
{\partial\over \partial r}
\left( \frac{\mu \Omega}{\omega_p+\Omega -\kappa}
    {\partial g\over \partial r}\right)
  + \frac{4\omega_p\Sigma g}{\Omega} =
  \frac{2{\rm i}r^{2}}{GM_*}\int^{\infty}_{-\infty}
\rho\frac{\partial \Phi}{\partial z}dz \sim 
\frac{2{\rm i}r^{2}\Sigma}{GM_*}
\left(\frac{\partial \Phi}{\partial z}\right)_{z=0},
\label{inceq01}
\end{equation}
where \be \mu =\int^{\infty}_{-\infty}\rho z^2dz, \ee $g = r^2 \Omega
v_z'/(r^3\Omega^2) \equiv v_z'/(r\Omega)$ and terms of order
$\omega_p^2$ have been neglected.  Note that consistently with the
approximations used here, where possible, we have adopted Keplerian
rotation and taken $r^3\Omega^2=GM_*$ to be a constant in this
expression.  Thus $2g$ becomes the local inclination of the disc (the
factor of two arises from the form of the Fourier decomposition
(\ref{Fourier})). We remark that (\ref{inceq01}) may also be written
as \be {\partial\over \partial r}\left(\frac{\mu
    \Omega}{\omega_p+\Omega -\kappa} {\partial g\over \partial
    r}\right) +\frac{4\omega_p\Sigma g}{\Omega} = \frac{1}{\pi
  GM_*}\frac{dT}{dr},
\label{inceq1}\ee
where, for orbits with line of nodes coinciding with the $y$ axis as
considered below, $T$ is the torque due to the planet acting on the
disc in the radial interval $(r,\infty)$ in the $x$ direction.  Thus
$T \rightarrow 0$ as $r\rightarrow \infty.$

\subsection{Solution for the disc inclination}

We wish to solve (\ref{inceq1}) for the inclination of the disc
induced by a planet on an inclined orbit.  As the unforced problem has
solutions consisting of long wavelength bending waves propagating with
a speed that is a multiple of the sound speed (Nelson \& Papaloizou
1999), a complete solution requires knowledge of the complete
structure of the disc including boundary details which are beyond the
scope of the modelling in this paper.  To deal with this situation we
assume the disc extends to large radii and has a much larger angular
momentum content than the planet. We then look for a solution for
which the inclination response is localized away from large radii
having $g \rightarrow 0,$ and $dg/dr \rightarrow 0,$ for $r
\rightarrow \infty.$ These solutions are in fact regular as $r
\rightarrow 0$ for the power law discs we adopt here.  However, inner
boundary effects could possibly cause the excitation of a freely
propagating bending wave that should be added. But this should not
affect an estimate of the scale of the warping.  To simplify matters
further, we shall take $\kappa=\Omega$ as expected for a constant
aspect ratio disc, under the gravitational potential due to a central
point mass, and assume that $\omega_p/\Omega$ is much less than $H/r.$
The latter assumption, which is expected to lead to mild warping (eg
Larwood et al. 1996 and see below) enables us to neglect the term
$\propto g$ in equation (\ref{inceq1}) which can then be easily
integrated to give \be \frac{\mu \Omega}{\omega_p} {\partial
  g\over \partial r} = \frac{1}{\pi GM_*}T,
\label{inceq2}\ee
and accordingly
 \be
 g  =-\omega_p \int^{\infty}_{r} \frac{T}{\pi\mu\Omega GM_*}dr.
\label{inceq3}\ee
We make a rough estimate of $g$ as determined by (\ref{inceq3}) by
setting $\mu = \Sigma H^2$ and $T \sim \pi GM_p\Sigma r\sim
M_pr^2\omega_p\Omega,$ where $r$ is evaluated at a location where the planet
intersects the disc.  Then we estimate \be g \sim \frac{\omega_p M_p
  r^2}{\Omega M_* H^2} \sim \left( \frac{\omega_p^2r^2}{\Omega^2
    H^2}\right) \left(\frac{ M_p\Omega}{M_*\omega_p }\right).  \ee
This implies that the inclination range is in general small. The first
factor in brackets measures the square of the product of the
precession frequency and local sound crossing time which is expected
to be less than unity and so lead to only small warping (see
Papaloizou \& Terquem 1995, Larwood et al. 1996, Nelson \& Papaloizou
1999).  The second factor in brackets is expected to be the ratio of
the planet mass to the characteristic disc mass contained within a
length scale comparable to that of the orbit, which is expected to be
less than of order unity particularly for low mass planets.

We have evaluated the solution given by (\ref{inceq3}) for a planet in
a circular orbit of radius $R,$ inclined at $45$ degrees to the plane
of the disc and with line of nodes coinciding with the $y$ axis.  We
took $\Sigma \propto r^{-1/2}$ as in the main text and as the solution
scales with the inverse square of the disc aspect ratio, $H/r,$ that
is left as a parameter.  The left hand panel of Fig. \ref{figA1} shows
the inclination in units of $10^3(M_p/M_*)(0.05R/H(R))^2 \omega_p
R^{3/2}/\sqrt{GM_*},$ as a function of $r/R.$ The maximum value of
this is of order unity indicating that, even for a Jovian mass planet
in a disc with aspect ratio $0.05,$ the characteristic value of the
inclination range is of order $\omega_p R^{3/2}/\sqrt{GM_*}$ which is
expected to be of the order of the ratio of the mass of the disc to
that of the central star and thus a small quantity.  The right hand
panel of Fig. \ref{figA1} shows the cumulative torque $T(r/R)/T(0)$
measured in the sense of increasing inwards.  This indicates that the
torque falls off rapidly at large radii.

\section{Expression  of the gravitational force due to the disc
 in terms of  elliptic
  integrals}
\label{app:potential}

Here we develop expressions for the gravitational force per unit mass
exerted by a disc on a planet that may be passing through it in
terms of elliptic integrals. The resulting expressions require the
evaluation of two dimensional integrals with integrands that at worst
contain a logarithmic singularity which is readily managable
numerically.

The gravitational potential exerted by the disk at the location
$(r,z)$ of the planet is given by equation~(\ref{potential}) as
\be \Phi (r,z_p) = -G \int_{R_i}^{R_o} \int_{-H}^{H}
\int_{0}^{2\pi} \frac{\rho(r',z') r' dr' dz'd\phi' }{\sqrt{r^2+r'^2 -
    2rr' \cos \phi' +(z-z')^2}},
\label{potential2}  \ee
As the disc is axisymmetric, the gravitational force per unit mass
exerted by the disc on the planet has only a radial and a vertical
components, given by $-\partial \Phi / \partial r$ and
$-\partial \Phi / \partial z$, respectively.
To calculate these, we first note that $\Phi$ is solution of
Poisson's equation, which is given by
\be \frac{\partial^2 \Phi}{\partial r^2} + \frac{1}{r} \frac{\partial
  \Phi}{\partial r} + \frac{\partial^2 \Phi}{\partial z^2} - \frac{m^2
  \Phi}{r^2} = 4 \pi G \rho(r,z) ,
\label{Poisson}
\ee
\noindent with $m=0$.  In the general nonaxisymmetric case, when the
$\varphi$ dependence  of $\rho$ is through a factor $\exp({\rm
  i}\varphi),$ the azimuthal number $m$ is nonzero and we have
$\partial^2 \Phi / \partial \varphi^2 = -m^2\Phi$.  We now
differentiate equation~(\ref{Poisson}), with $m=0,$ with respect to
$r$ to obtain to \be \frac{\partial^2}{\partial r^2} \left(
  \frac{\partial \Phi}{\partial r} \right) + \frac{1}{r}
\frac{\partial}{\partial r} \left( \frac{\partial \Phi}{\partial r}
\right) - \frac{ 1}{r^2} \frac{\partial \Phi}{\partial r} +
\frac{\partial^2}{\partial z^2} \left( \frac{\partial \Phi}{\partial
    r} \right) = 4 \pi G \frac{\partial \rho}{\partial r} .
\label{PoissonFr}
\ee

\noindent This shows that $\partial \Phi / \partial r$ satisfies
Poisson's equation with $m=1$ and $\partial \rho / \partial r$ as
source term.  The solution can be written as

\be \frac{\partial \Phi}{\partial r} = -G \int_{R_i}^{R_o}
\int_{-H}^{H} \int_{0}^{2\pi} \frac{\frac{\partial \rho}{\partial
    r}(r',z') r' dr' \cos \phi' dz'  d\phi'}{\sqrt{r^2+r'^2 - 2rr'
    \cos \phi' +(z-z')^2}} + B_i + B_o.
\label{Fr0}  \ee

\noindent Here the domain of integration is the interior of the domain
containing the mass distribution and $B_i$ and $B_o$ are boundary
terms that have to be taken into account when the disc's mass density
is given by equation~(\ref{rho}), as in that case $\partial \rho
/ \partial r$ is infinite at the radial boundaries of the disc.  When
the disc's mass density is made continuous by multiplication by the
factor $f$ defined in equation~(\ref{rhof}), these boundary terms are
not included.

\noindent Expression~(\ref{Fr0}) can be recast in  the form:

\be \frac{\partial \Phi}{\partial r} = -G \int_{R_i}^{R_o}
\int_{-H}^{H} \frac{\partial \rho}{\partial r}(r',z') H_r(r,r',z-z') r'
dr' dz' + B_i + B_o ,
\label{Fr}  \ee

\noindent with

\be H_r(r,r',z-z') = \int_{0}^{2\pi} \frac{\cos \phi' d \phi'}
{\sqrt{r^2+r'^2 - 2rr' \cos \phi' +(z-z')^2}}.
\ee

\noindent We define $a^2=r^2+r'^2+(z-z')^2$, $b^2=2rr'$ and
$u=2b^2/(a^2+b^2)$.  It is then straightforward to show that:

\be H_r(r,r',z-z') = \frac{4 \sqrt{a^2 + b^2}}{b^2} \left[
  \frac{a^2}{a^2+b^2} K(u) - E(u) \right], \ee

\noindent where $K$ and $E$ are the elliptic integrals of the first
and second kind, defined as:

\begin{eqnarray}
K(m) & = & \int_0^{\pi/2} \frac{d \theta}{\sqrt{1-m \sin^2 \theta}} , \\
E(m) & = & \int_0^{\pi/2} \sqrt{1-m \sin^2 \theta} d \theta,
\end{eqnarray}

\noindent with $m<1$.

\noindent We calculate the boundary terms $B_i$ and $B_o$  by 
assuming that $\rho$ is continuous and  supposing
that $\rho$ increases from 0 to $\rho(R_i,z)$ over a distance $\Delta
r \rightarrow 0$ at the inner edge, and decreases from $\rho(R_o,z)$
to 0 over the same distance at the outer edge, i.e.:

\begin{eqnarray}
B_i & = & -G \int_{R_i - \Delta r}^{R_i} \int_{-H}^H \frac{\partial \rho}{\partial
  r}(r',z') H_r(r,r',z-z') r' dr' dz', \\
& = & -G \int_{-H(R_i)}^{H(R_i)} \rho(R_i,z') R_i H_r(r,R_i,z-z') dz'.
\end{eqnarray}

\noindent Similarly:

\begin{eqnarray}
B_o & = & -G \int_{R_o }^{R_o + \Delta r} \int_{-H}^H \frac{\partial \rho}{\partial
  r}(r',z') H_r(r,r',z-z') r' dr' dz', \\
& = & + G \int_{-H(R_o)}^{H(R_o)} \rho(R_o,z') R_o H_r(r,R_o,z-z') dz'.
\end{eqnarray}

We calculate $\partial \Phi / \partial z$ in a similar way by
differentiating Poisson's equation~(\ref{Poisson}), with $m=0,$ with
respect to $z$, which leads to:

\be \frac{\partial^2}{\partial r^2} \left( \frac{\partial
  \Phi}{\partial z} \right) 
+ \frac{1}{r} \frac{\partial}{\partial r}
\left( \frac{\partial \Phi}{\partial z} \right) 
+ \frac{\partial^2}{\partial z^2}
\left( \frac{\partial \Phi}{\partial z} \right) = 4 \pi G
\frac{\partial \rho}{\partial z} .
\label{PoissonFz}
\ee

\noindent This shows that $\partial \Phi / \partial z$ satisfies
Poisson's equation with $m=0$ and $\partial \rho / \partial z$ as
source term.  The solution can be written as

\be \frac{\partial \Phi}{\partial z} = -G \int_{R_i}^{R_o}
\int_{-H}^{H} \int_{0}^{2\pi} \frac{\frac{\partial \rho}{\partial
    z}(r',z') r' dr' d\phi' dz' }{\sqrt{r^2+r'^2 - 2rr'
    \cos \phi' +(z-z')^2}}.
\label{Fz0}  \ee

\noindent This expression can be recast in  the form

\be \frac{\partial \Phi}{\partial z} = -G \int_{R_i}^{R_o}
\int_{-H}^{H} \frac{\partial \rho}{\partial z}(r',z') H_z(r,r',z-z') r'
dr' dz',
\label{Fz}  \ee

\noindent with:

\be H_z(r,r',z-z') = \int_{0}^{2\pi} \frac{d \phi'}
{\sqrt{r^2+r'^2 - 2rr' \cos \phi' +(z-z')^2}},
\ee

\noindent or, in term of the elliptic function $K$:

\be H_z(r,r',z-z') = \frac{4}{\sqrt{a^2+b^2}} K(u). \ee

%
%

%
\label{lastpage}
\end{document}